\documentclass[12pt]{article}
\usepackage[utf8]{inputenc}

\usepackage{amsmath,amssymb,amsfonts,amscd,mathrsfs}
\usepackage{xcolor}
\definecolor{darkblue}{rgb}{0.1,0.1,.7}
\usepackage[colorlinks, linkcolor=darkblue, citecolor=darkblue, urlcolor=darkblue, linktocpage]{hyperref} 
\usepackage[square, comma, compress,numbers]{natbib}
\usepackage[]{graphicx}
\usepackage{geometry}
\usepackage[margin=10pt,font=small,labelfont=bf]{caption}
\usepackage{ifthen}

\usepackage{epstopdf}
\usepackage{setspace}
\usepackage{anysize}
\usepackage{enumitem}
\usepackage{amsthm}
\usepackage{multicol}

\newcommand{\mrm}[1]{{\mathrm #1}}

\newcommand{\diffop}[2]{\ifthenelse{\equal{#2}{1}}{\frac{\mrm{d}}{\mrm{d} #1}}{\frac{\mrm{d}^#2}{\mrm{d} #1^#2}}}

\newcommand{\cb}[2][\Delta]{g_{#1,#2}}
\newcommand{\cbh}[1]{h_{\Delta,#1}}

\newcommand{\reef}[1]{(\ref{#1})}
\def\eps{\epsilon}
\def\eps{\epsilon}
\newcommand{\beq}{\begin{equation}} 
\newcommand{\eeq}{\end{equation}}
 
\def\nn{\nonumber}

\def\calO {{\cal O}}

\def\half{{\textstyle\frac 12}}

\def\ge{\geqslant}
\def\le{\leqslant}

\def\leq{\leqslant}
\def\nn{\nonumber}

\def\eps{\epsilon}

\numberwithin{equation}{section}
\begin{document}

\vspace*{-.6in} \thispagestyle{empty}
\begin{flushright}
CERN PH-TH/2015-253\\
\end{flushright}
\vspace{1cm} {\Large
\begin{center}
{\bf Remarks on the Convergence Properties \\of the Conformal Block Expansion}\\
\end{center}}
\vspace{1cm}
\begin{center}
{\bf Slava Rychkov$^{a,b,c}$, Pierre Yvernay$^{d,a}$ }\\[2cm] 
{
\small
$^{a}$ CERN, Theory Department, Geneva, Switzerland\\
$^{b}$ Laboratoire de Physique Th\'{e}orique de l'\'{E}cole Normale Sup\'{e}rieure (LPTENS), Paris, France
\\
$^{c}$ Sorbonne Universit\'es, UPMC Univ Paris 06, Facult\'e de Physique, Paris, France 
\\
$^{d}$ ENSTA ParisTech, Palaiseau, France
\normalsize
}
\\
\end{center}

\vspace{4mm}

\begin{abstract}
We show how to refine conformal block expansion convergence estimates from hep-th/1208.6449. In doing so we find a novel explicit formula for the 3d conformal blocks on the real axis. 
\end{abstract}

\vspace{.2in}
\vspace{.3in}
\hspace{0.7cm} October 2015


\newpage

\section{Introduction and formulation of the problem}

Many interesting new results about conformal field theories (CFTs) in $d>2$ dimensions have been recently obtained using the bootstrap approach. In this method one uses the operator product expansion (OPE) associativity to constrain the CFT data (local operator dimensions and their OPE coefficients). Operationally, one expands the four point functions in conformal partial waves and imposes that expansions in different channels give the same result. 

The resulting ``conformal bootstrap equations" have a nice feature that they are totally mathematically well defined, expressing the agreement between convergent power series with nonempty and overlapping regions of convergence \cite{Pappadopulo:2012jk}.
In practice one would like to know how fast these series converge. This question was already studied in \cite{Pappadopulo:2012jk}, and here we will give it a fresh look, but first let us explain why this is important. For us the main interest in this question comes from the need to put on solid ground Gliozzi's approach to the bootstrap. 


Recall that there are currently two competing approaches to the numerical bootstrap. In the first one, known as the linear/semidefinite (LSD) programming \cite{Rattazzi:2008pe,Caracciolo:2009bx,Rattazzi:2010yc,Poland:2011ey,ElShowk:2012ht,ElShowk:2012hu,El-Showk:2014dwa,Kos:2014bka,Simmons-Duffin:2015qma}\footnote{We just cite a few papers where important development steps of the method were made.}, one actually does not truncate the expansion, or rather truncates it at such high dimension and spin of the exchanged operators that the truncation error is absolutely negligible. The obtained results take form of rigorous bounds on the space of CFT data. It is using this approach that most numerical results were obtained. Here we will only highlight the study of the 3d Ising model which yielded world's most precise estimates on its leading critical exponents \cite{ElShowk:2012ht,El-Showk:2014dwa,Kos:2014bka,Simmons-Duffin:2015qma}.

In the second approach, by Gliozzi and collaborators \cite{Gliozzi:2013ysa,Gliozzi:2014jsa,Gliozzi:2015qsa}, one does truncate the expansion pretty severely, keeping just a handful of low-dimension operators. One then completely neglects the truncation error and, expanding around the usual middle point to a finite order, gets a system of nonlinear equations to solve for the dimensions of the retained operators and their OPE coefficients. This method has several advantages over the LSD programming: it is more intuitive, not so heavy on the numerical side, and is applicable for both unitary and non-unitary theories. On the other hand it is not as systematic, lacking as of now a built-in mechanism to estimate the truncation-induced error. There are e.g.~small but noticeable differences between the 3d Ising critical exponents determined by using Gliozzi's approach \cite{Gliozzi:2014jsa}
and the LSD programming \cite{El-Showk:2014dwa,Simmons-Duffin:2015qma}. The LSD results are rigorous; they also agree with Monte Carlo (while being more precise), while Gliozzi's approach does not.  

We consider it an important open problem to find a modification of Gliozzi's approach which would keep its above-mentioned positive features, while allowing to estimate an error induced by the truncation. Having a good control over the rate of convergence of the conformal block expansion is a prerequisite for this task.

For concreteness and simplicity, in this paper we will study a conformal four point function of identical Hermitean primary scalar operators:
\begin{equation}
\langle\phi(x_1)\phi(x_2)\phi(x_3)\phi(x_4)\rangle = g(u,v)/(x_{12}^2 x^2_{34})^{\Delta_\phi}\,.
\label{eq:4pt}
\end{equation}
The function $g(u,v)$ depends on the usual conformally invariant cross ratios and can be expanded in conformal blocks of the exchanged primaries:
\beq
g(u,v)=\sum_\calO \lambda^2_\calO\, g_\calO(u,v)\,.
\label{eq:CBexp}
\eeq
We are interested in the convergence rate of this expansion. We will work in the Euclidean space although the Minkowski case is also interesting \cite{Fitzpatrick:2012yx,Komargodski:2012ek,Hartman:2015lfa}. We will consider here the unitary case when all OPE coefficients are real and hence $\lambda^2_\calO>0$.

The convergence question was already studied in \cite{Pappadopulo:2012jk}. To state their result we need some CFT kinematics.
It is standard to parametrize any four point function configuration in the Euclidean space in terms of a complex number $z\in\mathbb{C}\backslash(1,+\infty)$. This is done by mapping the four points to a plane, and then moving them within this plane to positions $0,z,1,\infty$. The relation between $u,v$ and $z$ is
\beq
u=|z|^2,\quad v=|1-z|^2\,.
\label{eq:uvz}
\eeq
It is very convenient to further map the range of $z$ to the unit disk by introducing the variable
\beq
\rho=z/(1+\sqrt{1-z})^2,\qquad |\rho|<1\,.
\label{eq:rhoz}
\eeq
This has the meaning to placing the points at the positions $\rho,-\rho,1,-1$. The inverse transformation is $z=4\rho/(1+\rho)^2$. From now on we switch from $u,v$ to use $\rho$ as our main conformally invariant parameter. If needed, one can go back to $u,v$ via \reef{eq:uvz}, \reef{eq:rhoz}.

The convergence bound proved in \cite{Pappadopulo:2012jk} states:\footnote{$a\approx b$ and $a\lesssim b$ $(x\rightarrow x_0)$ mean $\lim_{ x\rightarrow x_0}(a/b)=1$ and $\limsup_{ x\rightarrow x_0}(a/b)\leq 1$, respectively.}
\beq
\left|\sum_{\calO:\Delta_\calO>\Delta_*}\lambda^2_\calO\, g_\calO(\rho)\right|\lesssim
\frac{(2\Delta_*)^{4\Delta_\phi}}{\Gamma(4\Delta_\phi+1)}|\rho|^{\Delta_*}\,, \qquad \Delta_*\to\infty\,.
\label{eq:conv1}
\eeq
This should be read as follows. In the lhs we have the tail of the conformal block expansion corresponding to the primary operators of dimension above some cutoff dimension $\Delta_*$ (and all spins). This is the error we will make in the four point function if we drop all such operators. We now hold $\rho$ fixed and take the limit $\Delta_*\gg 1$.\footnote{More precisely, the bound \reef{eq:conv1} sets in for $\Delta_*\gg \Delta_\phi/(1-|\rho|)$ \cite{Pappadopulo:2012jk}.} Since any configuration corresponds to $|\rho|<1$, we see that for large $\Delta_*$ the tail becomes exponentially small, because of the factor $|\rho|^{\Delta_*}$. One configuration particularly important for the bootstrap analysis is $z=\half$, which gives $\rho=3-2\sqrt{2}\approx 0.17$. We see that the convergence at this point is quite fast.

Simple numerical experiments in gaussian CFTs where the conformal block expansion is exactly known can be used to check that the exponentially decreasing factor $|\rho|^{\Delta_*}$ in \reef{eq:conv1} is best possible. On the other hand, the same experiments indicate that the power of $\Delta_*$ in the prefactor is not optimal. Improving the prefactor will be the main goal of our paper.

\section{Review of \cite{Pappadopulo:2012jk}}
\label{sec:review}

That the prefactor in \reef{eq:conv1} is not optimal has a simple origin in the way that estimate was derived in \cite{Pappadopulo:2012jk}. Let's review the derivation, and then see how it can be improved.

\emph{Step 1}. One observes that the conformal blocks have an expansion of the form \cite{Hogervorst:2013sma}
\beq
g_\calO(\rho )=\sum_{\delta,j} f_{\delta, j} C_j(\cos\phi)  r^\delta \,,
\label{eq:CBrad}
\eeq
where in the lhs $\rho=r e^{i\phi}$, in the rhs $\delta,j$ are the dimensions and spins of the primary operator $\calO=\calO_{\Delta,l}$ and of its descendants. In particular $\delta=\Delta+n$, $n\in\mathbb{N}_0$, while $j$ ranges between $l\pm n$. The $C_j$ are Gegenbauer polynomials in $\cos\phi$. The only important thing for us is that they take their maximal value, normalized to 1, at $\phi=0$. Finally $f_{\delta,j}$ are relative coefficients of descendants with respect to that of the primary. Let's normalize the conformal block by setting the primary coefficient to one: $f_{\Delta,\ell}=1$. The rest of the coefficients are then fixed by conformal symmetry. It's important that they are all nonnegative:
\beq
f_{\delta,j}\ge 0\,.
\eeq
This condition is satisfied as long as the primary field is above the unitarity bound (which is true since we assume we have a unitary theory).

\emph{Step 2}. Two simple consequences of \reef{eq:CBrad} are :
\beq
|g_\calO(re^{i\phi})|\lesssim |g_\calO(r)|\,,\qquad g_\calO(r)>0\,.
\eeq
This means that it's enough to study convergence at real $\rho$, since for nonzero $\phi$ it will be only faster.

\emph{Step 3}. So from now on we specialize to the real axis. Consider two series representations of the same function $g(r)$: the conformal block expansion
\beq
g(r)=\sum_\calO \lambda_\calO^2\, g_\calO(r)
\label{eq:gr}
\eeq
and
\beq
g(r)=\sum_\delta p_\delta\, r^\delta\,.
\label{eq:2ndrep}
\eeq
To obtain the second series we simply plug in the series representation \reef{eq:CBrad} of each conformal block into \reef{eq:gr} and collect all powers of $r$ with their respective coefficients. So $p_\delta= \sum\lambda_\calO^2 f_{\delta,j}$ where the sum is over all descendants having the scaling dimension $\delta$ (no matter what $j$ and the primary are). Clearly $p_\delta\ge 0$. We will call a power series whose all coefficients are positive a \emph{coefficient-positive series}.

Now the tail of the first series is strictly smaller than the tail of the second series:
\beq
\sum_{\calO:\Delta_\calO>\Delta_*} \lambda_\calO^2\, g_\calO(r) < \sum_{\delta>\Delta_*} p_\delta\, r^\delta\,.
\label{eq:2tails}
\eeq
This is because the tail on the rhs contains all terms which are present on the lhs, and in addition it contains contributions 
of descendants of dimension $\delta>\Delta_*$ coming from primaries of dimension $\Delta_\calO\le \Delta_*$.

As a matter of fact, Ref.~\cite{Pappadopulo:2012jk} established the convergence estimate on the tail of the second series,
and used \reef{eq:2tails} to say that the first tail satisfies the same estimate. 
It is here that the prefactor optimality is lost.  Here we will try to do better, but first let us complete the review.

\emph{Step 4}. Notice that the full four point function with points positioned at $r,-r,1,-1$ has in the limit $r\to 1$ the asymptotics
\begin{equation}
\label{eq:4.collision}
\langle\phi(r)\phi(-r)\phi(1)\phi(-1)\rangle \approx {1}/{(1-r)^{4\Delta_\phi}}\,,
\end{equation}
which follows from applying the OPE for the two pairs of operators which become close to each other. Notice that as usual we normalize the operators with the unit two point function coefficient. Given \reef{eq:4pt}, this gives the asymptotics for $g(r)$:
\beq
g(r) \approx {2^{4\Delta_\phi}}/{(1-r)^{4\Delta_\phi}}\qquad(r\gg1)\,.
\label{eq:asg}
\eeq
Putting together this fact with the power series representation \reef{eq:2ndrep}, we find ourselves within the assumptions of the Hardy-Littlewood (HL) tauberian theorem, which establishes the asymptotics for the \emph{integrated} coefficients $p_\delta$:
\beq
P(E):=\sum_{\delta \le E} p_\delta 
\approx
 \frac{(2E)^{4\Delta_\phi}}{\Gamma(4\Delta_\phi+1)}, \quad E\rightarrow \infty\,.
\label{eq:HLres}
\eeq
A physicist would normally try to fit a powerlaw assumption about $p_\delta$ to the asymptotics \reef{eq:asg}, obtaining
\beq
p_E \approx \frac{2^{4\Delta_\phi}E^{4\Delta_\phi-1}}{\Gamma(4\Delta_\phi)}\,\qquad\text{(naive)},
\label{eq:naive}
\eeq
and then by integration \reef{eq:HLres}. Notice however that in a CFT $p_\delta$ is not a smooth function but a discrete sequence. It's not a totally obvious matter to show that the result \reef{eq:HLres} still holds under these circumstances, The assumption $p_\delta\ge 0$ is crucial. The naive argumentation gives a quick way to recover the answer but it's not a substitute to the proof. In fact it's known to fail completely for the subleading terms in the asymptotics. 
See the book \cite{Korevaar} for a thorough review of the HL theorem and its ramifications.

From \reef{eq:HLres}, the tail can be estimated as follows (denote $r=e^{-t}$, $t>0$):
\begin{align}
\sum_{\delta>\Delta_*} p_\delta\, e^{- t \delta} &= t \int_{\Delta_*}^\infty dE\, [P(E)-P(\Delta_*)] e^{-tE}\nn\\
&\le t \int_{\Delta_*}^\infty dE\,P(E) e^{-tE}\nn\\
&\approx t \int_{\Delta_*}^\infty dE\, \frac{(2E)^{4\Delta_\phi}}{\Gamma(4\Delta_\phi+1)} e^{-tE}
 \lesssim \frac{(2\Delta_*)^{4\Delta_\phi}}{\Gamma(4\Delta_\phi+1)}e^{-t\Delta_*}\,.
 \label{eq:chain}
\end{align}
Here we integrated by parts in the first line, then dropped the negative term $-P(\Delta_*)$, used the HL estimate on $P(E)$, and the asymptotic behavior of the incomplete gamma function to conclude. The last step requires $\Delta_* \gg \Delta_\phi / t$.

Putting together this estimate and \reef{eq:2tails} we obtain \reef{eq:conv1} for real $r<1$. By Step 2, the same bound is valid for all complex $\rho=r e^{i\phi}$.

It's instructive to compare \reef{eq:chain} with an estimate we would have obtained if we just approximated the sequence $p_\delta$ by the naive powerlaw \reef{eq:naive}:
\beq
\sum_{\delta>\Delta_*} p_\delta\, e^{- t \delta} \approx \int^\infty_{\Delta_*} dE\,p_E\, e^{- t E} \approx \frac{2^{4\Delta_\phi}\Delta_*^{4\Delta_\phi-1}}{t\,\Gamma(4\Delta_\phi)}\,
e^{-t\Delta_*}\,\,\qquad\text{(naive)}\,.
\eeq
The reason why this naive estimate has a better prefactor can be traced back to having dropped $-P(\Delta_*)$ in the chain of estimates leading to \reef{eq:chain}. If $P(E)$ is a nice powerlaw, there is a near cancelation between $P(E)$ and $-P(\Delta_*)$ for $E$ close to $\Delta_*$, which sharpens the bound.\footnote{We thank Petr Kravchuk and Hirosi Ooguri for discussions concerning this point.} To justify this cancelation in general, we would need some sort of subleading asymptotics for $P(E)$, and unfortunately this is not available; see the discussion after Eq.~(4.20) in \cite{Pappadopulo:2012jk}. So \reef{eq:chain} is the best we currently have in full generality.

\section{3d conformal blocks on the real axis}
\label{sec:3d}
As indicated in the previous section, the loss in the prefactor of the convergence rate estimate of \cite{Pappadopulo:2012jk} comes from treating the primaries and the descendants in the conformal blocks on equal footing. There are many terms in the rhs of \reef{eq:2tails} which are not present in the lhs. To improve the estimate we should think of a conformal block as a whole, instead of separating it into constituents. However, conformal blocks in general are complicated special functions, while the HL theorem is for sums of powers of $r$. While there are generalizations of the HL theorem valid for more general functions of $r$, here we will demonstrate a more simple-minded approach. 

We stumbled on the possibility of this approach while studying the conformal blocks in 3d. We will therefore start by presenting these results, which are of independent interest.
 
As shown in \cite{Hogervorst:2013kva}, conformal blocks on the real axis satisfy an ordinary differential equation. This is not obvious since as a function of complex $\rho$ they satisfy a partial differential equation. However there are in fact two PDEs: the well-known second order one coming from the quadratic Casimir, and in addition a fourth order one coming from the quartic Casimir. Taken together these two PDEs imply an ODE on the real axis.

The relevant ODE is obtained by specializing to the case of equal external operator dimensions by setting $P=S=0$ in (4.10a) of \cite{Hogervorst:2013kva}. It takes the form
\begin{equation}\label{eq:ODE}
\mathcal{D}_4\, \cb{l}(r) = 0
\end{equation}
with 
\begin{small}
\begin{align*}\mathcal{D}_4 &= 
(r -1)^3 (r +1)^4 r^4 \diffop{r}{4}+ 2(r -1)^2 (r+1)^3 r^3\{(2 \eps + 5) r^2 + 2 \eps - 1\}\diffop{r}{3} \\
	&-2(r -1)(r +1)^2 r^2 \left\{\left(c_2-(\eps + 4) (2 \eps + 3)\right) r^4 - 2 (2 \eps^2 + c_2 + 3 \eps - 5)r^2 - 2\eps^2 + c_2 + \eps\right\}\diffop{r}{2}\\ 
	&-2(r +1)r \left\{(2\eps + 3) (c_2 -2(\eps + 1)) r^6 + (4 (-2\eps^2 +\eps + 3) - c_2 (2\eps + 5))r^4 \right.\\
	&\quad \left.+  \left(c_2 +2(\eps-1)\right)(1-2\eps) r^2 + (1 + 2\eps) c_2\right\}\diffop{r}{1}\\
	&+ (1-r) \{(2 (2\eps + 1) c_2 - c_4)r^6 + 2 (-c_4 + 2 (2\eps + 1) c_2)r^5 + (c_4 + 2 (6\eps - 1) c_2)r^4\\
	& \quad+4r^3 (c_4 + 2 c_2 (2\eps - 1)) + (c_4 +2 c_2 (6\eps - 1)) r^2 + 2(2 c_2 (2\eps + 1) - c_4)r + 2c_2(2\eps + 1) -c_4\}\,.
\end{align*}
\end{small} 
Here $\eps= d/2-1$ while $c_2$ and $c_4$ are the quadratic and quartic Casimir eigenvalues expressed in terms of the primary dimension and spin:
\begin{equation}
c_2 =  \half [l(l+2\eps)+\Delta(\Delta-2-2\eps)], \quad c_4 = l(l+2\eps)(\Delta-1)(\Delta-1-2\eps).
\end{equation}

It turns out that the above equation can be solved in $d=3$ by a judicious substitution. To guess the substitution, consider first the conformal blocks of the ``leading twist" operators for which
\beq
\Delta=l+2\eps,\qquad l=0,1,2\ldots
\eeq
For $l\ge 1$ this is the minimal dimension allowed by the unitarity bound, and the corresponding operators are the conserved currents. In the $z$ variable, these conformal blocks can be inferred from Eq.~(2.32) in \cite{Hogervorst:2013sma} (see also (A.6) in \cite{Hogervorst:2013kva}): 
\beq
g_{l+2\eps,l}(z)= z^{l+2\eps} {}_2F_1(l+\eps,l+2\eps;2l+2\eps;z)\qquad(z<1\text{ real})\,.
\eeq
This is valid for all $d$. For $d=3$ we can use the hypergeometric identity (see \cite{Bateman1}, 2.1.5)
\beq
z^{a}{}_2F_1(a,a+\half;b;z)=(4r)^a{}_2F_1(2a,2a-b+1;b;r)\,.
\eeq
Adjusting the normalization constant, we obtain an astonishingly simple formula
\beq
g_{l+1,l}(r)=r^{l+1}/(1-r^2)\qquad(d=3)\,.
\label{eq:lead}
\eeq

Consider now the case of $d=3$ and general $\Delta=l+\tau$, where $\tau$ is the twist. We propose the following substitution:
\beq
g_{\Delta,l}(r)=r^\Delta h_{\Delta,l}(x)/(1-x),\qquad x=r^2.
\eeq
The variable $x$ is natural since it is known that for equal external dimensions the conformal block stripped off from the leading $r^\Delta$ has an expansion in even powers of $r$ \cite{Hogervorst:2013sma}. Separating $1/(1-x)$ is then motivated by \reef{eq:lead}.

Specializing in \reef{eq:ODE} to $d=3$ and rewriting it in terms of $h_{\Delta,l}(x)$ we obtain
\begin{equation}\label{eq:ODE_simple}
\tilde{\mathcal{D}}\cbh{l}(x) = 0\,,
\end{equation}
where
\begin{align}
\tilde{\mathcal{D}} = ~& 8(x-1) x^3 \diffop{x}{4}+ 8\{(2 \Delta +5)x-2 \Delta -3\}x^2\diffop{x}{3}\nn\\
 & +2\{[4 \Delta ^2+2 \Delta  (\tau +10)-\tau ^2+\tau +15]x-4 \Delta ^2-2 \Delta  (\tau +4)+\tau ^2-\tau -3\}x\diffop{x}{2}\nn\\
& +\{(2 \Delta +1) [2 \Delta  (\tau +2)+(1-\tau ) \tau ]x
+\tau  (-4 \Delta ^2+2 \Delta  \tau -\tau+1)\}\diffop{x}{1}\nn \\
&+(\Delta -1) (\tau -1) (2 \Delta -\tau ).
\end{align}
We recognize the differential equation for the generalized hypergeometric function ${}_4F_3$, so we can write the solution in closed form:
\begin{equation}
\label{eq:3D_CB}
h_{\Delta,l}(x) =
 {}_4F_3\left[\frac{1}{2},\frac{\tau-1}{2},\Delta-\frac{\tau}{2},\Delta-1;\Delta-\frac{1}{2},\Delta+\frac{1-\tau}{2},\frac{\tau}{2};x\right]\,\quad (d=3).
\end{equation}
This explicit formula came as a surprise to us since normally conformal blocks in odd dimensions are hard to deal with. In even dimensions there are explicit formulas for any complex $z$ \cite{Dolan:2000ut,Dolan:2003hv,Dolan:2011dv} from which one can specialize to real $z$ but the results (see section \ref{sec:synthesis}) look still more complicated than \reef{eq:3D_CB}. For general $d$ on the real axis there are formulas in terms of ${}{}_3F_2$'s \cite{ElShowk:2012ht,Hogervorst:2013kva}, but they contain several terms unless $l=0,1$.

\section{Improved convergence estimate}

Using the results from the previous section, we will now improve the convergence rate estimate in $d=3$. Consider the function $\hat g(r)$ defined by removing from $g(r)$ contributions of all scalars of dimension $1/2<\Delta<1$, if there are any such scalars exchanged. The rationale for doing this will be clear below. For now notice that $g(r)$ and $\hat g(r)$ have the same asymptotics:
\beq
\hat g(r)\approx g(r)\approx 2^{4\Delta_\phi}/(1-r)^{4\Delta\phi}\qquad (r\to 1)\,.
\eeq
This is because the subtracted conformal blocks have $(1-r)^{-1}$ singularity and cannot change the asymptotics (notice that ${}_4F_3$ stays finite as $r\to1$).
 
The key idea is to apply the argument from section \ref{sec:review} not to the function $g(r)$ but to the function
\beq
\tilde g(r)=(1-r^2) \hat g(r)\,.
\eeq
which has the accordingly modified asymptotics
\beq
\tilde g(r)\approx 2^{4\Delta_\phi+1}/(1-r)^{4\Delta_\phi-1}\qquad (r\to 1).
\eeq
On the other hand the same function admits a representation
\beq
\tilde g(r)=\sum_{\calO:\Delta_\calO\ge 1}\lambda_\calO^2\, \tilde g_{\calO}(r),\qquad
\tilde g_{\Delta,l}(r)=r^\Delta h_{\Delta,l}(r^2)\,.
\eeq
From the explicit formula for \reef{eq:3D_CB} we see that $h_{\Delta,l}(x)$ is a coefficient-positive series in $x$. In fact, the coefficients are all positive as long as $\tau\ge 1$, which must be true by unitarity for $l\ge 1$. If there were scalars in the interval $1/2<\tau<1$, their $h_{\Delta,l}(x)$ functions would have negative coefficients, but we removed all such scalars by hand.

We can now run Steps 3,4 of the section \ref{sec:review} argument verbatim and get the following tail estimate:
\beq
\sum_{\calO:\Delta_\calO\ge \Delta_*}\lambda_\calO^2\, \tilde g_{\calO}(r) \lesssim
\frac{2^{4\Delta_\phi+1} \Delta_*^{4\Delta_\phi-1}}{\Gamma(4\Delta_\phi)}\,r^{\Delta_*}
\eeq
Dividing by $(1-r^2)$ we get the corresponding estimate in terms of conformal blocks:
\beq
\sum_{\calO:\Delta_\calO\ge \Delta_*}\lambda_\calO^2\, g_{\calO}(r) \lesssim
\frac{2^{4\Delta_\phi+1} \Delta_*^{4\Delta_\phi-1}}{\Gamma(4\Delta_\phi)}\,r^{\Delta_*}/(1-r^2)
\label{eq:final}
\eeq
This improves the previous result \reef{eq:conv1} by one power of $\Delta_*$ in the prefactor.\footnote{Notice however that the constant coefficient is a factor of $8\Delta_\phi$ larger, so the new estimate is nominally better than the old one only for $\Delta_*>8\Delta_\phi$ (assuming that $r^2\ll 1$).}

\section{Synthesis and conclusions}
\label{sec:synthesis}

The method used in the previous section can be stated synthetically as follows. Considers a representation of the conformal blocks factoring out a power of $1/(1-r^2)$:
\beq
g_\calO(r)=\tilde g_\calO(r)/(1-r^2)^\gamma\,,\qquad \gamma\ge 0.
\label{eq:rep1}
\eeq
The $\calO$-independent parameter $\gamma$ should be chosen so that the functions $\tilde g_\calO(r)$, like conformal blocks themselves, are coefficient-positive series in $r$. Or at least this last property should hold for all operators $\calO$ of dimension larger than some fixed dimension $\Delta_0$.  Under these assumptions one obtains the convergence rate estimate
\beq
\sum_{\calO:\Delta_\calO\ge \Delta_*}\lambda_\calO^2 g_{\calO}(r) \lesssim
\frac{2^{4\Delta_\phi+\gamma} \Delta_*^{4\Delta_\phi-\gamma}}{\Gamma(4\Delta_\phi+1-\gamma)}\,r^{\Delta_*}/(1-r^2)^\gamma\,,\qquad \Delta_*\gg \Delta_\phi/(1-r),
\eeq

In $d=3$ we had $\Delta_0=1$, $\gamma=1$. Moreover from the ``leading twist" conformal blocks \reef{eq:lead} we know that this is best possible.

What about the even dimensions? Notice that a necessary condition to have the representation \reef{eq:rep1} is that the conformal block should grow as $r\to1$ at least as fast as $1/(1-r)^\gamma$. From the explicit expressions from \cite{Dolan:2000ut,Dolan:2003hv} it is easy to get that the $d=2,4$ conformal blocks behave in this limit as 
\beq
g_{\Delta,l}(r)\sim 
\begin{cases}
\log \frac{1}{1-r},&d=2\,,\\
\frac{1}{(1-r)^2}\log \frac{1}{1-r},&d=4\,.
\end{cases}
\eeq
It follows that in $d=2$ the value $\gamma=0$ is optimal and we cannot improve the convergence rate estimate.

In $d=4$ the necessary condition says that we could go at most to $\gamma=2$. To determine the actual value of $\gamma$ we need a more detailed analysis. The exact expression for the $d=4$ blocks on the real axis which follows from \cite{Dolan:2000ut,Dolan:2003hv} is
\begin{align}
\label{eq:rep4d}
g^{d=4}_{\Delta,l}(r)&=\frac{r^\Delta}{1-r^2} \{f_{\Delta+l}(r^2)f_{\Delta-l-2}(r^2)\\
&\quad+\frac{2 r^2}{l+1}
[f'_{\Delta+l}(r^2)f_{\Delta-l-2}(r^2)-f_{\Delta+l}(r^2)f'_{\Delta-l-2}(r^2)]\}\,.\nn
\end{align}
It's easy to see from here that $\gamma=1$ is allowed. Indeed, both terms in braces are coefficient-positive series as long as $\Delta\ge l+2$ i.e.~above the spin $l$ unitarity bound. Moreover, having analyzed Eq.~\reef{eq:rep4d} analytically and numerically, it seems that the optimal value lies between 1 and 2, with $\gamma=3/2$ being our best bet. However we do not have a rigorous proof of this fact.

In conclusion, in this paper we presented a method which allows to improve convergence rate estimates for the conformal block expansions in various dimensions over the previous work \cite{Pappadopulo:2012jk}. We presented improved results in $d=3,4$, while in $d=2$ improvement turns out impossible. We believe that such studies will be important for the future implementations of truncation schemes in the conformal bootstrap.

{\bf Note added:} An interesting analysis of the OPE coefficient asymptotics and convergence properties of the conformal block expansion has just appeared in \cite{Ooguri}, especially concerning the limit when the dimensions of the external operator become large. There are nice similarities between our formulas for conformal blocks with powers of $1/(1-r^2)$ taken out, and the results of their appendix A.

\section*{Acknowledgements}

S.R is grateful to Sheer El-Showk, Hirosi Ooguri, Hugh Osborn, Miguel Paulos, Marco Serone and Alessandro Vichi for the useful discussions. S.R.'s research was partly supported by the National Centre of Competence in Research SwissMAP funded by the Swiss National Science Foundation. P.Y.'s research was supported by the CERN Summer Student program; he's grateful to the CERN Theory Department for the hospitality.

\addcontentsline{toc}{part}{References}
\bibliographystyle{utphys}
\bibliography{3Dconvergence-bibli}


\end{document}